\documentstyle[twoside,fleqn,psfig,espcrc2]{article}
\begin{document}
\renewcommand{\refname}{\normalsize\bf References}
\title{Effect of Dephasing on Charge-Counting Statistics in Chaotic
       Cavities} 
\author{Ya. M. Blanter%
        \address{D\'epartement de Physique Th\'eorique, Universit\'e
         de Gen\`eve, CH-1211 Gen\`eve 4, Switzerland\\[2mm]
        $^{\rm b}$Laboratoire de Physique Math\'ematique (CNRS UMR
         5825), Universit\'e 
         Montpellier II, Place E.~Bataillon, 34095 Montpellier,
         France}$^{,{\rm b}}$,\ %
	\refstepcounter{address}%
	H. Schomerus
        \address{Instituut-Lorentz, Universiteit Leiden, P.O.\ Box
         9506, 2300 RA Leiden, The Netherlands}, and
        C. W. J. Beenakker$^{\rm c}$}

\begin{abstract}
\hrule
\mbox{}\\[-0.2cm]

\noindent{\bf Abstract}\\
We calculate the cumulants of the charge transmitted through a chaotic
cavity in the limit that the two openings have a large number of
scattering channels. The shot noise, which is the second cumulant, is
known to be insensitive to dephasing in this limit. Unexpectedly, the
fourth and higher cumulants are found to depend on dephasing: A
semiclassical theory and a quantum mechanical model with strong
dephasing give a different result than a fully phase-coherent quantum
mechanical theory.
\\[0.2cm]
{\em PACS}: 05.60.-k, 03.65.Sq, 73.23.Ad, 73.50.Td
\\[0.1cm]
{\em Keywords}: Shot noise; Fluctuations; Quantum statistics;
Transport processes; Quantum dots
\\
\hrule
\end{abstract}
\maketitle

\section{Introduction}

The theory of shot noise in mesoscopic systems has developed along two
parallel lines, one fully quantum mechanical, the other semiclassical
(see \cite{Jong97,BB99} for reviews). The semiclassical method treats
the dynamics of the electrons according to classical mechanics, but
includes quantum statistical effects following from the Pauli
exclusion principle. The fully quantum mechanical method includes
interference effects that are ignored in the semiclassical
approach. Both methods are expected to
give identical results in the limit that the
Fermi wavelength goes to zero, or equivalently, in the limit that the
number of scattering channels $N$ goes to infinity,  corresponding to
a conductance large compared to the conductance quantum $e^2/h$. 

For example, in a diffusive conductor the shot-noise power $S\to
\frac{2}{3} e \langle I\rangle$ in the limit $N\to\infty$
(with $\langle I\rangle$ the mean current), and this result has been
obtained both quantum mechanically
(using random-matrix theory \cite{wires} or Green functions
\cite{Altshuler,Nazarov94,Nazarov95}) and semiclassically (using the
Boltzmann-Langevin equation \cite{Nagaev}).
Another example of the correspondence principle between quantum
mechanics and semiclassics in the large-$N$  limit is the shot noise
of a double-barrier tunneling diode
\cite{diodes1,diodes2,diodes3,Jong96}. 

The shot-noise power is the second cumulant of the charge
transmitted through the conductor in a certain time. For the complete
counting statistics, one needs to know the higher cumulants as well.
The only existing semiclassical calculation of higher cumulants was
done for the double-barrier tunneling diode \cite{Jong96}, and was
found to be in complete agreement with the quantum mechanical
theory \cite{counting1,counting2,counting3,LLY,counting4}
in the large-$N$ limit.

In this paper we present a semiclassical theory for the counting
statistics of charge transported through a chaotic cavity, using the
`minimal correlation' approach developed recently by 
Sukhorukov and one of the authors \cite{BS99}. Much to our surprise,
we do not recover the quantum mechanical results in the large-$N$
limit: the fourth and higher cumulants differ. We need to introduce
dephasing into the quantum mechanical theory (by means of a voltage
probe \cite{BIBM,Jong96a,Langen97}) to obtain agreement with
semiclassics. 

\begin{figure}
\centerline{\psfig{figure=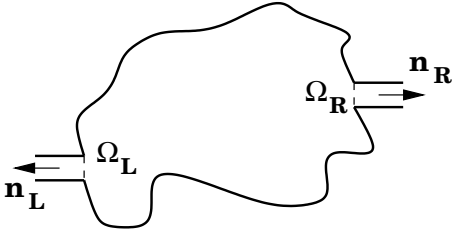,width=6.cm}}
\caption{Two-terminal chaotic cavity. The left and right contacts have
cross-section $\Omega_L$, $\Omega_R$, and outward normals ${\bf n}_L$,
${\bf n}_R$.}
\label{fig:1}
\end{figure}

\section{Transmitted charge}

We consider a  chaotic cavity (or quantum dot)
connected to two equilibrium reservoirs by ballistic contacts
(Fig.\ \ref{fig:1}). (Because our results do not depend on the
dimensionality of the cavity we
focus on the two-dimensional situation, which simplifies notation.)
A voltage $V$ is applied between the reservoirs. 
The reservoirs are described by equilibrium Fermi functions,
which at zero temperature (the case of interest in this paper) have
the form $f_L (E) = \theta(eV-E)$ and $f_R(E) = \theta(-E)$. The
cross-sections of the contacts are denoted by $\Omega_L$ and
$\Omega_R$. Since we want to exploit the chaotic dynamics of the
electron motion, we assume that there is no direct transmission
between the contacts. In addition, the contacts are narrow as compared
to the circumference of the cavity, but are classical in the sense
that the number $N_L$, $N_R$ of propagating modes through each contact
is large.

We are interested in the statistics of the charge transmitted
through any of the contacts in a given time interval $t$,
\begin{equation} \label{trcharge}
Q(t) = \int_0^t I(t')\,dt',
\end{equation}
where $I(t)=\langle I \rangle+\delta I(t)$
is the fluctuating current. We take the limit $t\to \infty$.
For definiteness,
we consider the current through the left contact, with the
convention that positive current is directed towards the sample.
In the long-time limit the charge transmitted through the right
contact is the same, due to current conservation.

The average transmitted charge gives the average current
$\langle I\rangle = \lim_{t \to \infty} t^{-1}\langle Q(t) \rangle$,
and the variance of the transmitted charge gives the zero-frequency
shot-noise power, 
\begin{eqnarray} 
S&=& 2\int_{-\infty}^{\infty}\langle \delta I(0)\delta I(t)\rangle\,dt
\nonumber
\\
&=&
2\lim_{t\to\infty}t^{-1}(
\langle Q^2 \rangle - \langle Q \rangle^2).
\end{eqnarray}
Higher cumulants of $Q(t)$
determine higher-order
correlation functions of the current. 

Classically, the current is expressed in terms of the distribution
function $f({\bf r}, {\bf n}, E, t)$, with ${\bf n}$ the
direction of momentum. The absolute value of the
momentum $p_F$ may be regarded as a constant.
The current is given by
\begin{eqnarray} \label{curdef1}
I_L (t) & = & -\frac{ep_F}{2\pi\hbar^2} \int_{\Omega_L} dy_L \int dE
\nonumber \\
&&{} \times  \int d{\bf n} \left( {\bf n}_L \cdot {\bf n} \right) f(x,
t), 
\end{eqnarray}
where ${\bf n}_L$ is the outward normal to the surface of the cavity
at the location of the left contact (Fig.\ \ref{fig:1}), $y_L$ is a
point at the cross-section of the contact,
and we have abbreviated $x \equiv \{ {\bf r}, {\bf n}, E \}$. 
The angular integral is normalized to unity, $\int d{\bf n} = 1$. 

Electrons incident from the left reservoir (${\bf n}_L \cdot {\bf n} <
0$) are described by the non-fluctuating distribution
function $f_L (E)$. (We ignore thermal fluctuations.) The
distribution function of the electrons with ${\bf n}_L  \cdot{\bf n} >
0$ carries information about the chaotic dynamics inside the
cavity. On average, it only depends on energy, since the coordinate
and angular dependence averages out due to the multiple scattering
from the surface of the cavity. This average distribution function
inside the cavity is readily found from the condition that the current
in each energy layer is conserved \cite{BS99},
\begin{equation} \label{avfun1}
\langle f \rangle = \frac{N_L f_L + N_R f_R}{N_L + N_R}.
\end{equation}
The number of modes $N_L$, $N_R$ is given by 
$N_{L,R} = p_F W_{L,R}/\pi\hbar$, with $W_{L,R}$ the width of the
contacts.
The mean
transmitted charge (\ref{trcharge}) is thus  
\begin{equation} \label{trchargeav}
\langle Q \rangle = \langle I_L \rangle t = \frac{e^2Vt}{h}
\frac{N_L N_R}{N_L + N_R}.
\end{equation}
This result could also be obtained by series addition of the ballistic
conductances $N_Le^2/h$ and $N_R e^2/h$ of the two contacts.

\section{Fluctuations of the distribution function}

Fluctuations of the transmitted charge at zero temperature are
entirely determined by the fluctuations of the distribution function
inside the cavity.
Following Ref.\
\cite{BS99}, we assume that the fluctuations of the distribution
function $\delta f \equiv f - \langle f \rangle$ inside the cavity
may be decomposed into two parts,
\begin{equation} \label{fluct1}
\delta f(x,t) = \delta \tilde f(x,t) + \delta f_C (E, t).
\end{equation}
The function $\delta \tilde f$ describes the purely ballistic
motion and obeys the equation 
\begin{equation} \label{Boltz}
\left( \partial_t + v_F {\bf n} \cdot \nabla \right) \delta \tilde
f(x,t) = 0, 
\end{equation}
with $v_F$ the Fermi velocity. This equation is supplemented by
the expression for the equal-time correlator, 
\begin{eqnarray} \label{fluct2}
\langle \delta \tilde f (x, t) \delta \tilde f (x', t) \rangle = 
\delta(x - x') \langle f \rangle \left( 1 - \langle f \rangle \right),
\end{eqnarray}
where 
\begin{equation}
\delta(x - x') = \nu^{-1} \delta({\bf r} - {\bf
r}') \delta({\bf n} - {\bf n}') \delta(E - E'),
\end{equation}
and $\nu = m/2\pi\hbar^2$ is the density of states. Eq.\
(\ref{fluct2}) has the same form as in equilibrium,
although the function $\langle f \rangle$ describes the
non-equilibrium state in the cavity.
The reason is that the only source of
noise at equilibrium as well as in the case of a deterministic chaotic
cavity is the partial occupation of states. The uniform
fluctuating term $\delta f_C$ accounts for the fact that at long times
the motion inside the cavity is not purely ballistic, and ensures the
current conservation at any time in any energy
interval.

The fluctuations in the current due to the term $\delta\tilde f$ are
given by
\begin{eqnarray} \label{curdef2}
 \delta \tilde I_{L} (t) &=& -\frac{ep_F}{2\pi\hbar^2}
\int_{\Omega_{L}} dy_{L} \int dE \nonumber \\
& & {}\times \int\limits_{({\bf n}_{L} \cdot {\bf n}) > 0} d{\bf n}\,
({\bf n}_{L} \cdot {\bf n}) \delta \tilde f(x, t),
\end{eqnarray}
and similarly for $\delta \tilde I_{R}(t)$.
The condition that the current at any time in any energy
interval is conserved can be used to
eliminate the fluctuation in the current due to the
term $\delta f_C$.
The total fluctuation of the current is then
given by
\begin{equation} \label{curdef3}
\delta I_L (t) =  \frac{N_R \delta \tilde I_L (t) - N_L \delta \tilde
I_R (t)}{N_L + N_R}.
\end{equation}

Eqs.\ (\ref{fluct1}) -- (\ref{fluct2}) constitute the minimal
correlation approach, which was shown in Ref. \cite{BS99} to
agree in the limit $N_L$, $N_R\gg 1$
with the quantum mechanical expression for shot noise in
a multi-terminal chaotic cavity. We seek to extend this approach to
higher cumulants of the transmitted charge. Therefore we have to
specify all the cumulants of the function $\delta \tilde f$ at equal
time moments. Generalizing Eq.\ (\ref{fluct2}), we write 
\begin{eqnarray} 
\lefteqn{\langle\langle \delta \tilde f(x_1, t) \delta \tilde f(x_2,
t) \cdots \delta \tilde f(x_k, t) \rangle \rangle = \delta (x_1 - x_2)
} 
\nonumber
\\
\label{fluct3} 
&& {} \times \delta (x_1 - x_3) \cdots \delta (x_1 - x_k)
\langle\langle f^k (x,t) \rangle\rangle, 
\end{eqnarray}
where $\langle\langle \cdots \rangle\rangle$ denotes the cumulant.
The cumulant $\langle\langle f^k (x,t)
\rangle\rangle$ is calculated later on; it depends only on $E$,
 since moments of
$f(x)$ are determined by $\langle f \rangle$. 
Eq.\ (\ref{fluct3}) is equivalent to Eq.\ (\ref{fluct2}) for $k=2$,
since $\langle\langle f^2\rangle\rangle=\langle f \rangle(1-\langle
f\rangle)$ (see below). 

Taking into account Eq.\ (\ref{Boltz}) and using Eq.\ (\ref{fluct3})
as the initial condition, we obtain the expression for the cumulant of
$\delta \tilde f$ at arbitrary times,
\begin{eqnarray} \label{fluct4}
\lefteqn{ \langle\langle \delta \tilde f(x_1, t_1) \delta \tilde f
(x_2,t_2) \cdots \delta \tilde f(x_k, t_k) \rangle\rangle \nonumber} \\
& = & \nu^{1-k} \delta(E_1 - E_2) \cdots \delta(E_{k-1}-E_k)
\nonumber \\
& &{}\times  \delta({\bf n}_1 - {\bf n}_2) \cdots \delta({\bf n}_{k-1} -
{\bf n}_k) \nonumber \\
& &{}\times  \delta\left[ {\bf r}_1 -
{\bf r}_2 - v_F {\bf n}_1 (t_1 - t_2) \right] \cdots \delta\left[ {\bf
r}_{k-1} \right. \nonumber \\
& & \left. \qquad-{\bf r}_k - v_F {\bf n}_1 (t_{k-1} - t_k) \right]
\langle\langle f^k \rangle\rangle. 
\end{eqnarray}

\section{Semiclassical cumulants}

Eqs.\ (\ref{curdef1}),
(\ref{curdef2}), (\ref{curdef3}), and (\ref{fluct4}) may now be used
to calculate the cumulants of the transmitted charge. We note that all
cumulants of the type $\langle\langle \delta \tilde I_{\alpha} (t_1)
\cdots \delta \tilde I_{\eta} (t_k) \rangle\rangle$ vanish due to the
combination of delta functions in Eq. (\ref{fluct4}), unless all the
subscripts $\alpha, \dots, \eta$ are equal to $L$ or all are equal to
$R$. The two non-zero cumulants are 
\begin{eqnarray} \label{curcum1}
\lefteqn{ \langle\langle \delta \tilde I_{\alpha} (t_1) \delta \tilde
I_{\alpha} (t_2)  \cdots \delta \tilde I_{\alpha} (t_k)\rangle\rangle =
(-1)^k \frac{e^kN_{\alpha}}{h} } \nonumber \\
& & {}\times \int dE \langle\langle f^k (E) \rangle\rangle \delta(t_1
- t_2) \cdots \delta(t_{k-1} - t_k), \nonumber \\
& & \alpha = L,R, \ \ \ k \ge 2. 
\end{eqnarray} 
Consequently, for the cumulant of the transmitted charge
we obtain
\begin{eqnarray} \label{charcum1}
\langle\langle Q^k \rangle\rangle & = & \frac{e^k t}{h}
\frac{N_L^k N_R + (-1)^k N_R^k N_L}{(N_L + N_R)^k} \nonumber \\
& &{}\times  \int dE\, \langle\langle f^k(E) \rangle\rangle. 
\end{eqnarray}

To complete the calculation, we must compute
the cumulant $\langle\langle f^k \rangle\rangle$. The fluctuating
distribution function assumes only the values $0$ and
$1$, thus $\langle f^k \rangle  = \langle f \rangle$ for $k \ge
1$. The characteristic function
\begin{equation} \label{charfun1}
\chi (p) = \left\langle \exp(pf) \right\rangle = 1 + \langle f \rangle
\left[ \exp(p) - 1 \right] 
\end{equation}
generates the cumulants as coefficients in  a series expansion,
\begin{equation} \label{charfun2}
\ln \chi (p) = \sum_{k=1}^{\infty} \frac{p^k}{k!} \langle\langle
f^k \rangle\rangle.
\end{equation}

Substituting the average distribution function (\ref{avfun1}),
integrating over energy, and taking into account
Eq.\ (\ref{trchargeav}), we obtain the expression for the cumulants of
the transmitted charge ($k \ge 2$), 
\begin{eqnarray} 
\langle\langle Q^k \rangle\rangle = e^{k-1} \langle Q \rangle
S_k \frac{N_L^{k-1} + (-1)^k N_R^{k-1}}{(N_L + N_R)^{k-1}}, 
\nonumber
\\
\label{charcum3}
\end{eqnarray}
where the coefficients $S_k$ are defined as
\begin{eqnarray} \label{coef1}
\ln \left[ N_R + N_L \exp(p) \right] = \sum_{k=0}^{\infty}
\frac{p^k}{k!} S_k.
\end{eqnarray}

Expression (\ref{charcum3})
may be simplified in the symmetric case $N_L = N_R$, when
\begin{equation} \label{coef2}
S_{2l} = \frac{2^{2l} - 1}{2l} B_{2l},\quad S_{2l+1}=0,\quad
l \ge 1,
\end{equation}
where $B_n$ are Bernoulli numbers. In this case we have
(for $l\ge 1$)
\begin{eqnarray}
\langle\langle Q^{2l} \rangle\rangle & = &e^{2l-1} \langle Q
\rangle \frac{2^{2l} -1}{l 2^{2l-1}} B_{2l}, 
\nonumber \\
\langle\langle Q^{2l+1} \rangle\rangle & = & 0.
\label{charcum4}
\end{eqnarray}
In particular, $\langle\langle Q^2 \rangle\rangle = e\langle
Q \rangle/4$, which is the $1/4$-shot noise suppression in
a symmetric chaotic cavity \cite{JPB94}. The next
non-vanishing cumulant is negative, $\langle\langle Q^4 \rangle\rangle
= -e^3 \langle Q \rangle/32$.  

For non-equivalent contacts Eq.\ (\ref{charcum3}) yields
\begin{equation} \label{noiscl}
\langle\langle Q^2 \rangle\rangle = e \langle Q \rangle \frac{N_L
N_R}{(N_L + N_R)^2}, 
\end{equation}
in agreement with Refs.\ \cite{Nazarov95,BeenRMP}.
The next two cumulants are
\begin{eqnarray}
\lefteqn{\langle\langle Q^3\rangle\rangle =-
e^2\langle Q\rangle
\frac{N_LN_R(N_L-N_R)^2}{(N_L+N_R)^4}
,}
\label{thirdcl} \\
\lefteqn{\langle\langle Q^4\rangle\rangle =
e^3 \langle Q\rangle N_LN_R}
\nonumber \\
&&{}\times
\frac{(N_L^2-4N_LN_R+N_R^2)(N_L^3+N_R^3)}{(N_L+N_R)^7}
. \label{fourthcl} 
\end{eqnarray}

\section{Phase-coherent quantum mechanical cumulants}

We produced the result (\ref{charcum3}) by a semiclassical method,
generalizing the minimal correlation approach to higher cumulants. Let
us compare it to the phase-coherent quantum
mechanical theory in the large-$N$ limit.

We use the relationship derived by Lee, Levitov, and Yakovets
\cite{LLY} between the characteristic function of the transmitted
charge and the transmission eigenvalues $T_j$, 
\begin{eqnarray}
\lefteqn{\langle \exp(Qp/e) \rangle = \prod_j \left[ 1 + T_j \left(
\exp(p) - 1 \right) \right]^{eVt/h}.  }
\nonumber \\
\label{cfcharge1}
\end{eqnarray}
This is the expression for a particular cavity, so we still have to
perform an ensemble average. The transmission eigenvalues have density
$\rho(T)$  in the ensemble of chaotic cavities.
The
ensemble-averaged cumulants of the transmitted charge follow from 
\begin{eqnarray}
\lefteqn{\sum_{k=1}^\infty \frac{p^k}{k!}e^{-k}
\langle\langle Q^k\rangle\rangle}
\nonumber
\\
&&=\frac{eVt}{h}\int_0^TdT\,\rho(T)\ln\left[1 + T\left(
\exp(p) - 1 \right) \right]
.
\nonumber
\\
{}&&
\label{eq:qrho}
\end{eqnarray}

The density of transmission
eigenvalues for $N_L$, $N_R\gg1$ has the form \cite{Nazarov95,BeenRMP}
\begin{eqnarray}
\lefteqn{\rho(T)=\frac{\sqrt{N_L N_R}}{\pi T} \left(\frac{T}{1-T}-
\frac{(N_L-N_R)^2}{4N_LN_R}\right)^{1/2}
.}
\nonumber\\
\label{eq:drho}
\end{eqnarray}
It vanishes for $T\le [1+4N_LN_R (N_L-N_R)^{-2}]^{-1}$.

In the symmetric case $N_L=N_R=N$ one has simply
\cite{JPB94,Baranger94} 
\begin{equation}
\rho(T)=\frac{N}{\pi}\frac{1}{\sqrt{T(1-T)}}
,
\end{equation}
and hence
\begin{equation}
\sum_{k=1}^\infty \frac{p^k}{k!}e^{1-k}\langle\langle
Q^k\rangle\rangle =4\langle Q\rangle \ln[{\textstyle \frac{1}{2}}+
{\textstyle \frac{1}{2}}\exp(p/2)]
,
\end{equation}
which is equivalent to Eqs.\ (\ref{charcum3}) and (\ref{coef2}). Thus,
for the symmetric case the results for the charge counting statistics,
obtained semiclassically and quantum mechanically, are identical. 

This correspondence does not extend to the more general case $N_L\neq
N_R$. From Eqs.\ (\ref{eq:qrho}) and (\ref{eq:drho}) 
we reproduce the semiclassical expressions (\ref{noiscl}) and
(\ref{thirdcl}) for $\langle\langle Q^2 \rangle\rangle$ and
$\langle\langle Q^3 \rangle\rangle$. However, the fourth cumulant,
\begin{eqnarray}
\lefteqn{ \langle\langle Q^4\rangle\rangle = e^3 \langle Q\rangle
N_LN_R (N_L+N_R)^{-6} } \nonumber \\
\lefteqn{\quad{}\times (N_L^4 -8N_L^3N_R + 12 N_L^2 N_R^2 - 8N_L N_R^3
+ N_R^4),}
 \nonumber   
\\
\label{fourthqm} 
\end{eqnarray} 
is different from Eq.\ (\ref{fourthcl}).

\section{Effect of dephasing on the quantum mechanical cumulants}

\begin{figure}
\centerline{\psfig{figure=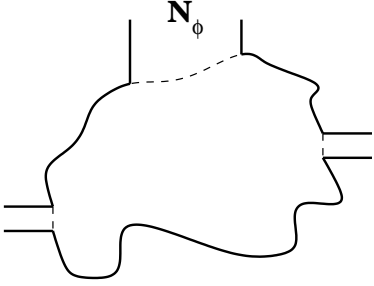,width=6.cm}}
\caption{Chaotic cavity with a fictitious lead (supporting $N_\phi$
propagating modes) that models strong uniform dephasing.} 
\label{fig:2}
\end{figure}

Apparently, the difference between Eqs.\ (\ref{fourthcl}) and
(\ref{fourthqm}) is due to the lack of phase coherence in the
semiclassical minimal correlation approach. To strengthen this
explanation, we incorporate into the quantum mechanical theory a
dephasing mechanism via an additional fictitious lead attached to
the cavity (see Fig.\ \ref{fig:2}).
Following Refs.\ \cite{Jong96a} and \cite{Langen97},
we assume that this lead has a
(fluctuating) distribution function which is determined from the
condition that no current flows into the reservoir at every energy and
every instant of time. An electron absorbed by this lead is
immediately re-injected back at the same energy, without any
memory of the phase. Thus,
this fictitious lead introduces dephasing but
not inelastic scattering. For homogeneous and complete
dephasing, we take the number of scattering channels $N_{\phi}$ in the
dephasing lead very large, $N_{\phi} \gg N_L$, $N_R$.

Our starting point is the expression for the current operator in the
lead $\alpha$ ($\alpha = L,R, \phi$) in terms of
creation/annihilation operators \cite{diodes1,BB99},
\begin{eqnarray} \label{curqm}
\lefteqn{
\hat I_{\alpha} (t)  =  \frac{e}{h} \sum_{\gamma\delta} \sum_{mn}
\int dE \int dE' \exp \left( \frac{i(E - E')t}{\hbar} \right)
}
\nonumber \\
&&{} \times  \hat a^{\dagger}_{\gamma m} (E) A^{mn}_{\gamma\delta}
(\alpha, E, E') \hat a_{\delta n} (E').  
\end{eqnarray}
The sum over $\gamma$ and $\delta$ is over the leads $L$, $R$, and
$\phi$, the sum over $m$, $n$ is over the mode indices. The current
matrix $A$ is expressed in terms of the scattering matrices $s$, 
\begin{eqnarray}
\lefteqn{
A^{mn}_{\gamma\delta} (\alpha, E, E') = \delta_{mn}
\delta_{\alpha\gamma} \delta_{\alpha\delta} -
s^{\dagger mn}_{\alpha\gamma} (E) s_{\alpha\delta}^{nm} (E'),
}
\nonumber \\
\label{curmatr}
\end{eqnarray}
and the creation and annihilation operators obey the fermion
anticommutation relation 
\begin{eqnarray} \label{comrule1}
\hat a^{\dagger}_{\alpha m} (E) \hat a_{\beta n} (E') & + & \hat
a_{\beta n} (E') \hat a^{\dagger}_{\alpha m} (E) \nonumber \\
& = & \delta_{\alpha\beta} \delta_{mn} \delta(E-E'). 
\end{eqnarray}

The expectation value
\begin{equation} \label{avoper1}
\langle a^{\dagger}_{\alpha m} (E) \hat a_{\beta n} (E')
\rangle = \delta_{\alpha\beta} \delta_{mn} \delta(E-E')
f_{\alpha} (E)
\end{equation}
is given by the average distribution function $f_{\alpha}(E)$ in
reservoir $\alpha$. From the condition that the average (energy
resolved) current $\langle I_{\phi} (E) \rangle$ through the dephasing
lead vanishes at every energy, one finds the average
distribution function 
\begin{equation} \label{avfunqm1}
f_{\phi} (E) = \frac{N_L f_L + N_R f_R}{N_L + N_R},
\end{equation}
which is identical to the average semiclassical distribution function
(\ref{avfun1}) inside the cavity. 

The distribution function of the dephasing lead fluctuates in
time. These fluctuations, which can also be found from the condition
that the current through this lead vanishes, modify the
fluctuations of the current at the leads  $L$ and $R$ \cite{Langen97},
\begin{eqnarray} \label{curfl1}
\Delta \hat I_L (t) = \delta \hat I_L (t) + \frac{N_L}{N_L + N_R}
\delta \hat I_{\phi} (t), \nonumber \\
\Delta \hat I_R (t) = \delta \hat I_R (t) + \frac{N_R}{N_L + N_R}
\delta \hat I_{\phi} (t),
\end{eqnarray}
where the intrinsic fluctuations $\delta \hat I_{\alpha} (t)=\hat
I_{\alpha} (t)-\langle \hat I_{\alpha} (t) \rangle$ are
described by Eq.\ (\ref{curqm}). 

We now calculate the fourth cumulant of the charge transmitted through
the left lead, 
\begin{equation} \label{cumqm1}
\langle\langle Q^4 \rangle\rangle = \langle\langle
\left( \int_0^t dt' \Delta \hat I_L (t') \right)^4
\rangle\rangle    
.
\end{equation} 
Explicitly, we have 
\begin{eqnarray} \label{cumqm2}
\lefteqn{\langle\langle Q^4 \rangle\rangle = \Xi_{LLLL} +
\frac{4N_L}{N_L + N_R} \Xi_{LLL\phi} }\nonumber \\
&&{}+ \frac{6N_L^2}{(N_L + N_R)^2} \Xi_{LL\phi\phi} +
\frac{4N_L^3}{(N_L + N_R)^3} \Xi_{L\phi\phi\phi} \nonumber \\
&&{}+ \frac{N_L^4}{(N_L + N_R)^4} \Xi_{\phi\phi\phi\phi},     
\end{eqnarray}
where we have defined the cumulant
\begin{equation}
\Xi_{\alpha_1\alpha_2\alpha_3\alpha_4}=
\langle\langle \prod_{i=1}^4 \int dt_i\,\hat I_{\alpha_i}
(t_i)\rangle\rangle 
\end{equation} 
in terms of the current operators (\ref{curqm}).
It remains to calculate the cumulant of eight
creation/annihilation operators, by
taking into account all possible pairings between the operators $\hat
a^{\dagger}_{\gamma_i m_i}$ and $\hat a_{\delta_i n_i}$ ($i=1$, 2, 3, 4)
which couple {\em all} the four current
operators together.
For $N_{\phi} \gg N_L, N_R \gg 1$ the leading contribution
to the cumulant comes from the terms with $\gamma_i =\delta_i =
\phi$. Summing all possible pairings, we obtain 
\begin{eqnarray} 
\lefteqn{
\Xi_{\alpha_1\alpha_2\alpha_3\alpha_4} = \frac{e^4t}{h} {\rm Tr} \
\left[ A_{\phi\phi} (\alpha_1) \cdots A_{\phi\phi} (\alpha_4) \right]
\int dE }
\nonumber \\ 
\lefteqn{\quad{}\times f_{\phi} \left( 1 - f_{\phi} \right) \left[
f_{\phi}^2 - 4f_{\phi} \left( 1 - f_{\phi} \right) + \left( 1 - f_{\phi}
\right)^2 \right],}
\nonumber \\
\end{eqnarray}
where the trace is taken over the mode indices. In the same
leading order, we neglect all traces of the type ${\rm Tr}\
[s^{\dagger}_{\alpha\phi} \cdots s_{\eta\phi}]$ ($\alpha \dots 
\eta = L, R$), unless all indices $\alpha \dots \eta$ are the
same. In this case 
\begin{equation}
{\rm Tr} \left[ s^{\dagger}_{L\phi} s_{L\phi} \right]^4 = N_L, \ \ \
{\rm Tr} \left[ s^{\dagger}_{R\phi} s_{R\phi} \right]^4 = N_R.
\end{equation}
Summing all contributions in Eq.\ (\ref{cumqm1}), we
recover the result (\ref{fourthcl}) of the semiclassical theory.

\section{Conclusions}

To summarize, we have studied the charge counting statistics in
chaotic cavities using three different approaches: (i) a fully
coherent quantum mechanical theory; (ii) a dephasing-lead model that
is also quantum mechanical but phenomenologically introduces uniform
and strong dephasing inside the cavity; and (iii) the semiclassical
minimal correlation approach. All three approaches give the same
results if the two openings in the cavity have the same (large) number
of scattering channels. The cumulants for this symmetric case are
given by Eq.\ (\ref{charcum4}). In the asymmetric case of 
two different openings the first three cumulants are also the same
in the three approaches, but the fourth cumulant is not: For approach
(i) it is given by Eq.\ (\ref{fourthqm}), for approaches (ii) and
(iii) by Eq.\ (\ref{fourthcl}). 

We conclude that the counting statistics of a chaotic cavity is
sensitive to dephasing even in the limit of a large number of
scattering channels. This is an unexpected conclusion since the shot
noise is not affected by dephasing. What is lacking is an
understanding in physical terms for why the high-order cumulants of
the transmitted charge respond differently to dephasing than the
low-order cumulants. 

We thank the Max-Planck-Institut f\"ur Physik Komplexer Systeme in
Dresden for hospitality and support. We acknowledge support by the
Swiss National Science Foundation via the Nanoscience program, by the
Dutch Science Foundation NWO/FOM, and by the TMR program of the
European Commission.

\end{document}